\font\caps=cmcsc10 at 12pt
\font\eightrm=cmr8 at 8pt
\documentclass{oxarticle}
\usepackage{amsmath, amssymb}
\usepackage{graphics, setspace}
\usepackage{epstopdf}
 \newcommand{\ET}{Exchange Transformation}

\newcommand{\ME}{Master Equation}

\newcommand{\cM}{{\cal M}}

\newcommand{\SG}{{Supergravity}}

\newcommand{\cG}{{\cal G}}

\newcommand{\cA}{{\cal A}}

\newcommand{\PB}{Master Equation}

\newcommand{\cL}{\cal L}
\newcommand{\bt}{\begin{tabular}{c}}
\newcommand{\et}{\end{tabular}}

\newcommand{\eb}{\ee\be } 
 
\newcommand{\bmat}{\lt ( \begin{array} }
\newcommand{\emat}{  \end{array} \rt )}

\newcommand{\oH}{{\ov H}}

\newcommand{\cP}{{\cal P}}

\newcommand{\ovD}{{\ov D}}

\newcommand{\oJ}{{\ov J}}

\newcommand{\oS}{{\ov S}}

\newcommand{\oG}{{\ov \G}}

\newcommand{\ED}{\end{document}}

\renewcommand{\a}{\alpha}

\renewcommand{\d}{\delta}

\newcommand{\z}{\zeta}
\newcommand{\h}{\eta}

\newcommand{\m}{\mu}
\newcommand{\n}{\nu}	
\newcommand{\x}{\xi}

\newcommand{\w}{\omega}
\newcommand{\G}{\Gamma}
\newcommand{\D}{\Delta}

\renewcommand{\P}{\Pi}	
\renewcommand{\S}{\Sigma}

\newcommand{\la}{\label}
\newcommand{\ci}{\cite}

\newcommand{\ds}{\documentstyle}	
\newcommand{\fr}{\frac}

\newcommand{\pa}{\partial}
\newcommand{\ov}{\overline}
\newcommand{\be}{\begin{equation}}
\newcommand{\ee}{\end{equation}}
\newcommand{\ba}{\begin{array}} 
\newcommand{\ea}{\end{array}}
\newcommand{\bea}{\begin{eqnarray}}
\newcommand{\eea}{\end{eqnarray}}
\newcommand{\ra}{\rightarrow}
\newcommand{\Ra}{\Rightarrow}

\newcommand{\lt}{\left}
\newcommand{\rt}{\right}

\newcommand{\Tr}{{\rm	Tr}	}

\newcommand{\ben}{\begin{enumerate}}
\newcommand{\een}{\end{enumerate}}
\newcommand{\bitem}{\begin{itemize}}

\newcounter{orange} 
\newcounter{apple} 
\addtocounter{apple}{1}

\newcounter{grape} 
\addtocounter{grape}{1}
\newcommand{\numberhere}{5274}
\newcommand{\articlenumber}{\numberhere{SuppSUSYjustYM}}

\proofmodefalse
\usepackage{color}

\newcommand{\mathsym}[1]{{}}
\newcommand{\unicode}[1]{{}}

\begin{document}
%\pagestyle{empty}

%\vspace*{1in}
 \begin{center}

{ \Huge The Basic Mechanism for Suppressed SUSY
\\[.5cm] }  

\vspace*{.1in}
%{    J. A. Dixon}
%

\renewcommand{\thefootnote}{\fnsymbol{footnote}}
%\footnotetext[1]{~here we have a footnote.}\renewcommand{\thefootnote}\arabicfootnote}}

{\caps John A. Dixon\footnote{jadixg@gmail.com}\\CAP\\Edmonton, Canada 
} \\[.5cm] 
%\vfill
\end{center}
\Large
\normalsize 
%\LARGE
%\Huge

 \begin{center}
 Abstract 
\end{center}

Suppressed SUSY is a new mechanism for `breaking SUSY'.   It requires \SG. It is independent of and very different from spontaneous or explicit  SUSY breaking. A recent paper illustrates some of its results. 
In this paper, the basic mechanism of Suppressed SUSY is explained in a simple SU(5) Yang Mills Theory, with a special set of Scalars.   Supersymmetry is not needed for this limited purpose.

\refstepcounter{orange}
{\bf \theorange}.\;
{\bf SUSY theories have Too Many Scalars:}  
In the last forty years, the Standard Model of strong weak and electromagnetic interactions has been repeatedly confirmed \ci{stdmodel}.  Supersymmetry (`SUSY')  does not seem relevant at all \ci{freepro,ferrarabook,superspace,WB,Buchbinder:1998qv,xerxes,west,Weinberg3,haber,buchmueller}. 
 Theoretically, 
the Standard Model starts off with one Higgs doublet, and this works very well indeed. 
However, SUSY starts off by requiring two Higgs doublets.  This is clearly a problem for SUSY, and it has remained a problem for forty years.

\refstepcounter{orange}
{\bf \theorange}.\;
{\bf Suppressed SUSY Suppresses Scalars:}  
 Suppressed SUSY starts at this point--it shows how to get back to one Higgs doublet, while keeping the power and uniqueness of the SUSY action.  It yields a way to SUPPRESS   the two Higgs doublets back into one\footnote{This suppression of real scalars from $8\ra 4$ is contained inside \ci{four,five}, which has $11 \ra 7$.  But since \SG\ is a necessary component of Suppressed SUSY, it is natural to want to look at a Grand Unified theory, with Planck scale masses, as in \ci{su5gust}. In that case 68 real scalar bosons get suppressed to 34. This suppression $68 \ra 34$ is the primary subject of the present paper.}.  Many consequences flow from that, as was seen recently in the simplest SU(5) Grand Unified Supergravity Theory \ci{su5gust}.
 
\refstepcounter{orange}
{\bf \theorange}.\;
{\bf Suppressed SUSY uses  \ET s  and the Master Equation:}  
 Suppressed SUSY \ci{su5gust} also gives rise to a new set of ideas about mass splitting and other predictions for SUSY.  
 The new mechanism works by profoundly changing the physics of the SUSY Action, while preserving all of the SUSY structure. This is accomplished by changing some of the Scalars of the theory into Sources rather than Fields.   This is put in place by writing down the Master Equation\ci{poissonbrak,Becchi:1975nq,zinnbook,taylor,Zinnarticle,becchiarticles1,becchiarticles2,BV,Weinberg2}
 in Supergravity\ci{freepro}, and using an `\ET'\ci{su5gust}.  
  For Supergravity this allows one  to shift some of the supersymmetry transformations onto the New Zinn Sources, thus avoiding many of the problems that arise for SUSY breaking, and from too many extra particles, while keeping much of the power of SUSY.
  
\refstepcounter{orange}
{\bf \theorange}.\;
{\bf The Master Equation has an important role in the physics:}  
    Up to now, the Master Equation has been largely regarded as an elegant way to incorporate symmetries into quantum field theory, with no real physical effect to be expected from it.  But in Suppressed SUSY, we see that it is also a way to generate theories that are physically very different from the starting theory, while keeping the same, very restrictive, algebraic symmetry.

\refstepcounter{orange}
{\bf \theorange}.\;
{\bf Suppressed Scalar Representations in Yang-Mills theory:}  
In this paper, we apply the basic ideas of Suppressed SUSY to a simple Yang--Mills model, without any SUSY. This  simplifies the exposition.   In fact, Supersymmetry is not needed to see what is happening.  However, the SUSY case is more interesting, and so, naturally enough, these ideas were noticed in that case first \cite{Dixon:2015cya,Dixon:2015fda,Dixon:2015bpa,four,five,su5gust}. In the present context, the result should probably be called `Suppressed Scalar Representations in Yang-Mills theory'.

\refstepcounter{orange}
{\bf \theorange}.\;
{\bf Contents of this Paper:}  This paper starts with a detailed short review and derivation of the Master Equation for a special Yang-Mills Theory with a special set of scalar representations (the `Old Theory').  Then in Paragraph \ref{exchangepara}, we implement the creation of a New Theory, from the Old Theory, by using a special  `\ET', which looks like a canonical transformation for the \ME, except that it interchanges some New and Old Fields with some New and Old Sources. The Master Equation
is a kind of Grassmann odd Poisson Bracket.   For a Poisson Bracket in classical mechanics, coordinates and momenta are equivalent under a canonical transformation \ci{goldstein,LandauLifmechanics}.  However Fields and Sources are NOT EQUIVALENT for the \ME\ in the Quantum Field Theory, because Fields are quantized, and Sources are not. The Quantum Field Theory arises from path integration of the Fields, leaving the Sources unintegrated.  If the Fields change, then the Quantum Field Theory changes too. In Paragraph  \ref{newFieldaction}, we set out the New  Action.  We derive the New Theory from the New Action, and then the New Master Equation in Paragraph \ref{newmasterfor1PI}.  A Summary of details, with the Old and New Field Actions, is in Paragraph \ref{summary}, and the issues arising from implementing the same \ET\ in Supergravity are discussed in the Conclusion in  Paragraph \ref{conclusion}.

\refstepcounter{orange}
{\bf \theorange}.\;
{\bf The Old Action with 68 Real Scalars:}  
  We take a simple model with $SU(5)$ Yang-Mills coupled to scalars, without any Supersymmetry whatsoever.  For the scalars we choose three special representations $H_L^i,H_{Ri},S^a$ because these are also important for Suppressed SUSY in \ci{su5gust}.  
 $H_L^i$ and $H_{Ri}$ are in the fundamental $5$ and $\ov 5$ representations.  $S^a$ is in a $24+24$ (complex adjoint) representation.

\refstepcounter{orange}
{\bf \theorange}.\;
{\bf Hermitian Matrices to Generate SU(5):}
As is well known, the simplest way to deal with the group SU(5), in the Yang-Mills context \ci
{ross,pran}, is to use the notation
\be
T^{ai}_j; i=1\cdots 5; j=1\cdots 5; a = 1\cdots 24.
\ee
for the 24 ($5 \times 5$)
hermitian generators of $SU(5)$ in the fundamental $5\times 5$ representation.  
Here are some results for Complex Conjugation $*$, Transposition $T$ and Adjoint $\dag$:   
\be
\lt ( T^{ai}_j \rt )^*
=\lt( T^{a i}_{j} \rt)^T
=  T^{a j}_{i}
;\;\lt ( T^{ai}_j \rt )^{\dag}
\equiv
\lt ( \lt ( T^{ai}_j \rt )^{*}\rt )^T
\equiv
\lt ( \lt ( T^{ai}_j \rt )^{T}\rt )^*
= T^{a i}_{j} 
\ee
The matrices $T^{ai}_j $ are also traceless. The  commutation relations are:
\be
 T^{ai}_j  
\d^j_i=T^{ai}_i  
=0; \d^i_i=5,
;\;  \lt [T^{a}, T^{b}\rt]   = i f^{abc}T^{c}    
;\; T^a = (T^a)^{\dag}
 \ee
 where we can choose to define $(T^a T^b)_i^j =  T^{aj}_k T^{bk}_i$.
Here we use the Kronecker $\d^i_j$ for the unit matrix and $f^{abc}$ is totally antisymmetric.

\refstepcounter{orange}
{\bf \theorange}.\;
{\bf Notation and Indices:}  
The multiplet $H_L^i $ transforms in the same way as the multiplet $\oH_{R}^{i}$.  Complex conjugation has the effect:
\be
\lt ( H_L^i \rt )^* =
\oH_{Li}
;\; \lt ( H_{Ri} \rt )^* =
\oH_{R}^{i}
 \ee
 Here we also have a complex adjoint $S^a$ representation.
This  $S^a$  is really two representations in the present context, since we could choose it to be real in this  Yang-Mills SU(5) context.  We can also choose the matrices to satisfy the following:
\be
S^{\;i}_{j}= S^a T^{ai}_j; \; T^{ai}_j T^{bj}_i = 2 \d^{ab} ; \;  S^{\;j}_{i}T^{ai}_j;  =2 S^a;
\;\Tr[S^2] = 2 S^a S^a;\; 
\ee
\be
\lt(S^a\rt)^* = \oS^a;\; \lt ( S^{i}_{j} \rt )^*= \oS^{j}_{i} = \oS^a T^{aj}_i;\;  \oS^{\;j}_{i}T^{ai}_j;  =2 \oS^a;\;\Tr[S \oS] = 2 S^a \oS^a
\ee

\refstepcounter{orange}
{\bf \theorange}.\;
{\bf Gauge Covariant Derivatives:}  
We define the gauge covariant derivative by:
\be
D_{ \m j}^{\;\;\;\;i} 
=
 \lt ( 
 \pa_{ \m} \d_{j}^{i} -
i V^a_{ \m} T_{j}^{ai}
 \rt )
\ee
In the above $V^a_{ \m}$ is the SU(5) Gauge Yang-Mills Field, and we absorb the coupling constant $g_5$, which is easy to restore.  
We raise the Lorentz index by:
\be
D_{ j}^{\m i} 
=
 \lt ( 
 \pa^{ \m} \d_{j}^{i} -
i V^{a  \m} T_{j}^{ai}
 \rt )
\ee
The Complex Conjugates are
\be
\lt ( D_{ \m j}^{i} \rt )^*
=
 \lt ( 
 \pa_{ \m} \d_{j}^{i} -
i V^a_{ \m} T_{j}^{ai}
 \rt )^*
=
 \pa_{ \m} \d_{i}^{j} +
i V^a_{ \m} T^{a j}_{i} 
\equiv \ovD_{ \m i}^{j} 
\ee
and we also need
\be
D^{ab}_{\m}
=
\lt ( 
 \pa_{ \m} \d^{ab}
 +
f^{acb} V^c_{ \m}  \rt )
\ee

Note that Complex Conjugation and Transposition raise and lower the indices $i,j=1 \cdots 5$.  But these have  no effect on the real adjoint index $a=1, \cdots 24$, or the Lorentz index $\m=0,1,2,3$. Because the matrices $ T^{ai}_j $ and $\d^j_i$ are Hermitian, Transposition is equivalent to Complex Conjugation for them. The horizontal (left-right) position of $i,j$ indices   and the  index $a$ on $ T^{ai}_j$ are not significant. The vertical (up-down) position of the index $a$ is not significant. But the up-down position of the index $\m$ is meaningful  because $\h_{\m\n}={\rm Diag(-1,1,1,1)}$ is not unity.

\refstepcounter{orange}
{\bf \theorange}.\;
{\bf The Old Zinn Source  Action:}  
\la{oldzinn}
As is usual for BRST analysis, one can start with a `Zinn Source 
Action'.  This consists of a set of expressions of the form
\be
\cA_{\rm Zinn} = \int d^4 x \lt \{\S^I \d {\cal B}_I
+ S^I \d {\cal F}_I \rt \}
\la{genzinn}
\ee
where $ {\cal B}_I$ are bosonic Fields or ghosts and ${\cal F}_I$ are fermionic Fields or ghosts.  The nilpotent BRS variations $\d$ of these are coupled to Zinn Sources $\S^I $ and $S^I$ so that $\cA$ is Grassmann even.  So, in (\ref{genzinn}),  $\d$ itself and the $\S^I$ are Grassmann odd, and the $S^I$ are Grassmann even.  We will start with  the following  
`Old Zinn' action:
\[
\cA_{\rm Old\; Zinn} = \int d^4 x \lt \{
i \G_{Ri}
 T^{ai}_{j } H_L^j
\w^a 
-i \ov \G_{R}^{i}
 T^{aj}_{i } \oH_{Lj}
\w^a 
-i \G_{L}^i
T^{aj}_{i } H_{R  j}
\w^a 
+i \ov \G_{Li}
T^{ai}_{j } \oH_{R}^{ j}
\w^a 
\rt.
\]
\be
\lt.
+
\G_{S}^a
f^{abc} S^b
\w^c 
+
\ov \G_{S}^a
f^{abc} \oS^b
\w^c 
+
\S^{a\m}
\lt ( 
 \pa_{ \m} \d^{ab}
 +
f^{acb} V^c_{ \m}  \rt )
 \w^b
- \fr{1}{2} M^{a}
f^{abc} \w^b
 \w^c
\rt\}
\la{oldzinneq}
\ee
In the above, $\w^a$ is the real Faddeev--Popov, Grassmann odd, Yang-Mills Ghost Field. 

\refstepcounter{orange}
{\bf \theorange}.\;
{\bf Variations $\d$ for the Old Fields:}  
\la{oldzinndelta}
Now consider the operator $\d$ defined by:
\be
\d H_{L}^i =   \fr{\d\cA_{\rm Old\; Zinn}}{\d \G_{Ri}}=    i 
 \w^a T^{ai}_{j } H_L^j
;\; 
\d \oH_{Li} = \fr{\d\cA_{\rm Old\; Zinn}}{\d \ov \G_{R}^{ i}}=-i 
 \w^a
T^{aj}_{i } \oH_{L  j}\ee
\be
\d H_{Ri} = \fr{\d\cA_{\rm Old\; Zinn}}{\d \G_{L}^{i}}=
-i  \w^a
T^{aj}_{i } H_{R  j}
;\; \d \oH_{R}^{i} = \fr{\d\cA_{\rm Old\; Zinn}}{\d \ov \G_{L i}}=
 i \w^a
T^{ai}_{j} \oH_{R}^{  j}
\ee
\be
\d S^a 
= \fr{\d\cA_{\rm Old\; Zinn}}{\d  \G_{S}^{ a}}
 =  f^{abc} S^b
\w^c 
;\; \d \oS^a = \fr{\d\cA_{\rm Old\; Zinn}}{\d  \ov\G_{S}^{ a}}
 =  f^{abc} \oS^b
\w^c 
\ee
\be
\d V_{\m}^a =  \fr{\d\cA_{\rm Old\; Zinn}}{\d  \S^{\m a}}
=  D^{ab}_{\m} \w^b
;\; \d \w^a =  \fr{\d\cA_{\rm Old\; Zinn}}{\d  M^{ a}}
=
-  \fr{1}{2}
f^{abc} \w^b
 \w^c
\ee

The above defined $\d$ is a real operator which satisfies the BRS equations of nilpotency:
\be
\d^2 =0
\ee
For example 
\be
\d^2 
H_{L}^i
=
    i 
\d \w^a T^{ai}_{j } H_L^j
-    i 
 \w^a T^{ai}_{j } \d H_L^j
=
 -   i 
 \fr{1}{2}
f^{abc} \w^b
 \w^c
 T^{ai}_{j } H_L^j
-    i 
 \w^a T^{ai}_{j } 
   i 
 \w^b T^{bj}_{k} H_L^k\ee
 
 \be
=
 -   i 
 \fr{1}{2}
f^{abc} \w^b
 \w^c
 T^{ai}_{k} H_L^k
+
\fr{1}{2} \w^a 
 \w^b
 \lt [
T^{a}
, T^{b}\rt ]^i_k H_L^k 
  =0
\ee

\refstepcounter{orange}
{\bf \theorange}.\;
{\bf Invariance of Scalar Kinetic Terms in Action: }  
\la{Higgskin}
We have the following gauge invariant kinetic terms in the Lagrangian:
\be
{\cL}_L = -
 D_{ \m j}^{i} 
H_{L}^{j} 
 \ovD_{i}^{\m k} 
\oH_{Lk} 
;\;{\cL}_R
 =
  -
 \ovD_{ \m j}^{i} 
H_{Ri} 
D_{k}^{\m j} 
\oH_{R}^{k} 
;\;  {\cL}_S 
=-
D^{ab}_{\m} S^b
D^{\m ad} \oS^d
\la{cLLold}
\ee

It is easy to show that
\be
\d   
\lt (D_{ \m i}^{j} 
H_{L}^{i} \rt )
=
i \w^b T^{bj}_{k}  D_{ \m i}^{k} 
H_{L}^i 
\ee
Here is the demonstration, to remind the reader how these things work:
\be
\d\lt(D_{ j}^{\m i} H_{L}^{i}\rt) 
=
 \pa_{ \m}\d  H_{L}^i 
 -
i\d V^{a  \m} T_{j}^{ai}
 H_{L}^j 
-
i V^{b  \m} T_{j}^{bi}
\d H_{L}^j 
\eb
=
 \pa_{ \m}  
 \lt (   i  \w^a T^{ai}_{j } H_L^j \rt)
 -
i \lt ( 
 \pa_{ \m} \w^a
 +
f^{acb} V^c_{ \m} \w^b \rt ) T_{j}^{ai}
 H_{L}^j 
-
i V^{b  \m} T_{j}^{bi}
\lt (   i  \w^a T^{aj}_{k } H_L^k \rt)
\eb
=
 \lt (   i  \w^a T^{ai}_{j }\pa_{ \m}  H_L^j \rt)
 -
i \lt ( 
f^{acb} V^c_{ \m} \w^b \rt ) T_{j}^{ai}
 H_{L}^j 
+
 V^{b  \m} \w^a 
 \lt ( T^a T^b + i f^{bac} T^c\rt )^i_kH_L^k 
=
i \w^b T^{bj}_{k}  D_{ \m i}^{k} 
H_{L}^i 
 \ee

and then complex conjugation implies that:
\be
\d   
\lt (\ov D_{ \m j}^{i} 
\ov H_{Li} \rt )
=
-i \w^b T^{bk}_{j}  \ov D_{ \m k}^{i} 
\ov H_{Li} 
\ee
and so
\be
\d {\cL}_L =0
\ee
Similarly:
\be
\d {\cL}_R =\d {\cL}_S =0
\ee

\refstepcounter{orange}
{\bf \theorange}.\;
{\bf Action to Start:}
\la{oldactionpara}  
The action $\cA_{\rm Start}$ for this Theory has been well known for many years.  Here is the Yang-Mills term:
\be
\cA_{\rm Old; \; Gauge} = \int d^4 x \lt \{
\fr{-1}{4} G^a_{\m\n} G^{a \m\n}
\rt \}
\la{oldgauge}
\ee
where $ G^a_{\m\n}= \pa_{\m} V^a_{\n}-\pa_{\n} V^a_{\m} + g^{abc} V^b_{\m} V^c_{\n} $ is the usual Yang Mills Field Strength. Then from paragraph \ref{Higgskin} we take the Scalar Kinetic terms, which contain 10+10+48=68 real scalar fields:
\be
\cA_{\rm Old; \; Higgs\;Kinetic\; Fields} = \int d^4 x \lt \{
{\cL}_L 
+
{\cL}_R+
{\cL}_S 
\rt \}
\la{oldkinetic}
\ee
Then there are a number of terms in the Higgs potential:
\be
\cA_{\rm Old; \; Higgs\; Potential} = \int d^4 x \lt \{
\cP\lt [H_L,H_R, S, \oH_L,\oH_R, \oS \rt]
\rt \}
\la{oldhigspot}
\ee
where this is a `gauge invariant' polynomial in the Scalar Fields, with no spacetime derivatives. It satisfies 
$\d \cP\lt [H_L,H_R, S, \oH_L,\oH_R, \oS \rt]=0
$. 

The kinetic term from (\ref{oldgauge}) is not invertible. To deal with this, we add the following to the definition of $\d$ in Paragraph \ref{oldzinndelta}:
\be
 \d \x^a =
  Z^a
;\; \d Z^a =0
\la{deltaz}
\ee
Then we define the following expression: 
\[
\cA_{\rm Old; \;Gauge\; Fixing}
= \int d^4 x \lt \{
\d \lt [ \x^a \lt ( \fr{\a}{2}Z^a+ \pa_{\m}V^{\m a} 
\rt) \rt ]+ U^{a}
\d \x^a 
\rt\}
\]\[
= \int d^4 x   \lt \{
Z^a \lt ( \fr{\a}{2}Z^a+ \pa_{\m}V^{\m a} 
\rt)
- \x^a 
 \lt ( 
 \pa_{\m}
D^{\m ab}\w^b
\rt )+ U^{a}
Z^a 
\rt\}
\]
\be
= \int d^4 x  \lt \{
\fr{\a}{2} 
\lt ( Z^a+ \fr{1}{\a} 
\lt(
 \pa_{\m}V^{\m a} + U^a \rt )
\rt)^2
-
\fr{1}{2 \a} 
\lt ( \pa_{\m}V^{\m a} + U^a 
\rt)^2
- \x^a 
 \lt ( 
 \pa_{\m}
D^{\m ab}\w^b
\rt )
\rt \}
\la{ggftostart}
\ee
In the above, $\x^a$ is  the real   Grassmann odd Yang-Mills Antighost Field and  $Z^a$ is a real  Grassmann even Gauge Fixing Auxiliary Field.  We have added a Zinn Source $U^a$ so that \be
 \d \x^a =
 \fr{\d \cA_{\rm Old;\; Gauge\; Fixing}}{\d  U^{ a}}
=  Z^a
\la{deltaz2}\ee

We will start our construction of the Old Quantum Field Theory with the 
Action:
\be
\cA_{\rm Start}= \int d^4 x {\cal L}_{\rm Start}= 
 (\ref{oldzinneq}) + 
(\ref{oldgauge})+
(\ref{oldkinetic}) +
(\ref{oldhigspot})
+(\ref{ggftostart})
\la{startaction}
\ee
It satisfies
\be
\d \cA_{\rm Start}=\d^2 =0
\ee
where $\d$ is as defined above in Paragraph \ref{oldzinndelta} and Equations (\ref{deltaz}) or (\ref{deltaz2}) or  (\ref{deltaz3}).
We can summarize these equations with the Master Equation form:
\be
\cM_{\rm Old}[\cA_{\rm Start}]   =0
\ee
where\footnote{The * indicates the complex conjugates of the three previous terms.  The last three terms are real.}
\be
\cM_{\rm Old}[{\cal X}] = \int d^4 x \lt \{ 
\fr{\d {\cal X}}{\d H_L^i}
\fr{\d {\cal X}}{\d \G_{Ri}}
+\fr{\d {\cal X}}{\d H_{Ri}}
\fr{\d {\cal X}}{\d \G_{L}^{i}}
+\fr{\d {\cal X}}{\d S^a}
\fr{\d {\cal X}}{\d \G_{S}^a} + *
+
\fr{\d {\cal X}}{\d V^a_{\m}}
\fr{\d {\cal X}}{\d \S^{a \m}}
+
\fr{\d {\cal X}}{\d M^a}
\fr{\d {\cal X}}{\d \w^a}
+
\fr{\d {\cal X}}{\d U^a}
\fr{\d {\cal X}}{\d \x^a}
\rt \}
\la{meold}
\ee

\refstepcounter{orange}
{\bf \theorange}.\;
{\bf The Generating Functional for Green's Functions:}  
First we define the Disconnected Green's Functional.  The path integral is integrated over all the Fields, Ghosts, Antighosts and the Gauge Auxiliary.  Define
\be
d ({\rm Old \;Fields})\;
=\P_{x,i,\m,a}    d H_L^i   d \oH_{Li}  d H_{Ri}   d \oH_{R}^i  d S^a d \oS^a  
d V_{\m}^a d \w^a d \x^a 
\la{doldfields}
\ee
and then take
\be
Z_{\rm Old}\lt [ j_{L}^{ k}, j_{R k},  j_{S}^{a} , j^a_{V, \m} ,
\z^a_{\w } , \z^a_{\x }; 
 \G_{Ri}
, \G_{L}^i
, \G_{S}^a
, \S^{a\m}
, M^{a}
,
U^a
\rt ]
=
\int   d ({\rm Old \;Fields})\; d Z^a 
e^{i \int d^4 y \lt \{
{\cL}_{\rm Start} +{\cL}_{\rm\;Old\; Field\; Sources}\rt \}}
\la{oldpathintfirst}\ee
In this path integral the integrand is defined by (\ref{startaction}). We   add the following Source terms for the Fields and ghosts and antighosts\footnote{The * indicates the complex conjugates of the three previous terms.  The last three terms are real. We do not add a source term for the field $Z^a$ because we can and will integrate it immediately.}.
\be
{\cL}_{\rm Old\;Field\; Sources} = \lt \{
j_{R k} H_L^k + j_{L}^{ k} H_{Rk}  + j_{S}^{a} S^a + *  +
j^a_{V,\m} V^{a \m}  -
\z^a_{\w } \w^{a}  -
\z^a_{\x } \x^{a} \rt \}
\la{oldFieldSources}\ee
The integration over $ d Z^a $ will be done in Paragraph (\ref{gaugfix}) below.

\refstepcounter{orange}
{\bf \theorange}.\;
{\bf Old Gauge Fixing and Ghost Terms:}  
 \la{gaugfix}
Our first step is to integrate the field $Z^a$.
We can shift the variable $Z^a\ra \lt ( Z^a- \fr{1}{\a} 
\lt(
 \pa_{\m}V^{\m a} + U^a \rt )
\rt)$ and then integrate the quadratic in $Z^a$ in Equation (\ref{ggftostart}), to get an irrelevant number. This leaves the following expression in place of Equation (\ref{ggftostart}):
\be
\cA_{\rm Old; \; Ghost,\; Gauge\; Fixing,\; and\;Antighost\; Zinn}
= \int d^4 x \lt \{
-
\fr{1}{2 \a} 
\lt ( \pa_{\m}V^{\m a} + U^a 
\rt)^2
- \x^a 
 \lt ( 
 \pa_{\m}
D^{\m ab}\w^b
\rt )
\rt \}
\la{laterggf}
\ee
and we note that, after the integration over $Z^a$, we get:
\be
 \d \x^a =  \fr{\d \cA_{\rm Start}}{\d  U^{ a}}
= 
   \fr{1}{ \a} 
\lt ( \pa_{\m}V^{\m a} + U^a 
\rt)
\la{deltaz3}
\ee

So now our action is slightly modified to the form
\be
\cA_{\rm Old}= \int d^4 x {\cal L}_{\rm Old}= 
 (\ref{oldzinneq}) + 
(\ref{oldgauge})+
(\ref{oldkinetic}) +
(\ref{oldhigspot})
+(\ref{laterggf})
\la{oldaction}
\ee
This  $\cA_{\rm Old}$ in (\ref{oldaction})
 still satisfies the Master Equation:
\be
\cM_{\rm Old}[\cA_{\rm Old}]=0
\la{oldmolda}
\ee

\refstepcounter{orange}
{\bf \theorange}.\;
{\bf A Change of Variables and the Old Jacobian Determinant}  
 \la{oldjacobian}
 Let us change variables in the integral, which is now:
\be
Z_{\rm Old}\lt [ j_{L}^{ k}, j_{R k},  j_{S}^{a} , j^a_{V, \m} ,
\z^a_{\w } , \z^a_{\x }; 
 \G_{Ri}
, \G_{L}^i
, \G_{S}^a
, \S^{a\m}
, M^{a}
,
U^a
\rt ]
=
\int   d ({\rm Old \;Fields})\;
e^{i \int d^4 y \lt \{
{\cL}_{\rm Old} +{\cL}_{\rm\;Old\; Field\; Sources}\rt \}}
\la{oldpathint}\ee

We take the following change of variables:
\be
{H'}_{L}^{i} =
H_{L}^i +  \varepsilon \fr{\d{\cA}_{\rm Old} }{\d \G_{Ri}}=
H_{L}^i + \varepsilon \d H_{L}^i \; {\rm etc.}
\la{hprime}
\ee
 and work our way through 
the Fields (\ref{doldfields}) using
(\ref{meold}). We do a similar exercise, with more detail, below in Paragraph \ref{newinv}. Here the parameter $\varepsilon$
is taken to be a Grassmann odd constant so that the Grassmann character of (\ref{hprime}) is even. 
What is the Jacobian here? 
We have:
\be
{H'}_{L}^i(x) =
H_{L}^i(x) + \varepsilon \d H_{L}^i (x) 
=
H_{L}^i(x) + \varepsilon  T^{ai}_j \w^a(x) H_{L}^j (x) 
\ee\
and we get
\be
\fr{\d {H'}_{L}^i(x)}{\d H_{L}^p(y)}
=
\d^4(x-y)\lt [ \d^i_p 
 + \varepsilon  T^{ai}_p \w^a(x) \rt ]
 \ee\
\be
\fr{\d {H'}_{L}^i(x)}{\d \w^b(y)}
=
\d^4(x-y)\lt [ - \varepsilon  T^{bi}_j H_{L}^j (x)  \rt ]
 \ee\
 
 Using 
 \be
 \rm Det \, Jac = e^{\rm Tr [ln\, Jac]}
\ee
this Jacobian can be seen to be 1 (following the methods  in \ci{zinnbook}).
Now, because of (\ref{oldmolda}), the variation of the Action is zero, and so we are left with just the variation of the term (\ref{oldFieldSources}).  Since a change of variables with unit Jacobian leaves the integral invariant, this variation must  be zero. 

\refstepcounter{orange}
{\bf \theorange}.\;
{\bf Derivation of the Old Master Equation for the 1PI Generating Functional:}  
 \la{oldmasterfor1PI}
So we see that:
\[
  \int   
d ({\rm Old \;Fields})\;
e^{i \int d^4 y \lt \{
{\cL}_{\rm Old} +{\cL}_{\rm\;Old\; Field\; Sources}\rt \}}
\]
\be
 \int d^4 x \; \lt \{ 
j_{R k} \d H_L^k + j_{L}^{ k} \d H_{Rk}  + j_{S}^{a} S\d ^a + *  +
j^a_{V,\m} \d V^{a \m}  +
\z^a_{\w } \d \w^{a}  +
\z^a_{\x } \d \x^{a} \rt \}
=0
\ee
Let us write this `expectation value'  in the form:
\be
 \int d^4 x \;  \lt < 
j_{R k} \fr{\d Z}{\d \G_L^k}
 + j_{L}^{ k}  \fr{\d Z}{\d \G_{Rk}}  + j_{S}^{a}  \fr{\d Z}{\d \G_S^a}+ *  +
j_{V,a \m}  \fr{\d Z}{\d \S_{a \m} }    +
\z^a_{\w }\fr{\d Z}{\d M_{a} }   +
\z^a_{\x } \fr{\d Z}{\d U^{a } }  \rt >_{\rm Old}=0
\la{wardfromshift}
\ee

Now define the Generating Functional $Z_C$ of Connected Green's Functions \ci{zinnbook}:
\be
Z_{\rm Old}= e^{i Z_{\rm C\;Old}}
\ee
and also define the Legendre Transform \ci{zinnbook} of the above to be  the Generating Functional $\cG_{1PI} $ of One Particle Irreducible Green's Functions:
\be
Z_{\rm C\;Old}= \cG_{\rm 1PI\; Old} + {\cL}_{\rm Old\;Field\; Sources}
\ee
where
\be
\fr{\d Z_{\rm C\;Old}}{\d H_L^i}= \fr{\d \cG_{\rm 1PI\; Old}}{\d H_L^i} +j_{Ri}=0
\ee
\be
\fr{\d Z_{\rm C\;Old}}{\d V_{\m}^a}= \fr{\d \cG_{\rm 1PI\; Old}}{\d V_{\m}^a} + j^a_{V,\m} =0
\ee
with similar definitions for the other Sources in expression (\ref{oldFieldSources}). 
Then (\ref{wardfromshift}) becomes 
\be
{\cM}_{\rm Old}\lt [\cG_{\rm 1PI\; Old} \rt ] =0
\ee

This is also true for the zeroth approximation to $\cG_{\rm 1PI\; Old} $, which
leads us back again to 
(\ref{oldmolda}).

\refstepcounter{orange}
{\bf \theorange}.\;
{\bf The Exchange Transformation:} 
\la{exchangepara}
The central theme of this paper is that we can now consider an `\ET' of the Master Equation 
(\ref{oldmolda}) above.  We generate this with 
the following Generating Functional:
\be
\cG =
 \int d^4 x \lt \{
\fr{1}{\sqrt{2}}
\lt ( 
H_L^i
+\ov H_R^i \rt)
\G_i
+
\fr{i}{\sqrt{2}}
\lt ( 
H_L^i
-\ov H_R^i \rt)
\h_i
+  S^a 
\fr{1}{\sqrt{2}}\lt(
 \G_K^a
 +  i \h_K^a
\rt )
\rt \} + *
\la{calG}
\ee

The New and Old Fields and Zinn Sources are as follows:
\ben
\item
The Old Scalar Fields $H_L^i, H_{Ri}, S^a$ in the Old Action (\ref{oldaction}) above, are conjugate to the Old Zinn Sources $\G_{Ri}, \G_{L}^i, \G_S^a$
\item
The New Scalar Zinn Sources  $J^i, \oJ_i,J^a$ in the New Action (\ref{newaction}) below,  are conjugate to New Antighosts $\h_i, \ov \h^i, \h^a$.  
\item
The New Scalar Fields are $H^i, \oH_i,K^a$ in the New Action (\ref{newaction}) below, are conjugate to the New Zinn Sources $\G_i, \ov \G^i, \G^a$
\een

The New Fields and the New $J$-type Zinn Sources are obtained as functions of the Old Fields as follows:
\[
H^i= \fr{\d \cG}{\d \G_i}=
\fr{1}{\sqrt{2}}
 \lt ( 
H_L^i
+\ov H_R^i \rt)
;
\;J^i= \fr{\d \cG}{\d \h_i}=
\fr{i}{\sqrt{2}}
 \lt ( 
H_L^i
-\ov H_R^i \rt)
;\;
\]\be
 K^a 
= \fr{\d \cG}{\d    \G_K^a}=
 \fr{1}{\sqrt{2}}\lt(
   S^a
+  \ov S^a 
\rt )
;\;
J_K^a 
= \fr{\d \cG}{\d    \h_K^a}=
 \fr{i}{\sqrt{2}}\lt(
   S^a
-  \ov S^a 
\rt )
 \ee
We can combine these to write the Old Fields in terms of the New Fields and  the New $J$-type Zinn Sources:
\[
H_L^i=
\fr{1}{\sqrt{2}}
\lt (H^i -i J^i\rt )
;\;
H_{Ri}=
\fr{1}{\sqrt{2}}
\lt (\oH_i -i \oJ_i\rt )
;\;
S^a 
=
 \fr{1}{\sqrt{2}}\lt(
K^a 
-i
J_K^a
\rt );
\]
\be 
\oH_{Li}=
\fr{1}{\sqrt{2}}
\lt (\oH_i +i \oJ_i\rt )
;\;
\ov H_R^i 
=
\fr{1}{\sqrt{2}}
\lt (H^i +i  J^i\rt )
;\;
\oS^a
=
 \fr{1}{\sqrt{2}}\lt(
K^a 
+i
J_K^a
\rt ).
\la{fsubs2}
 \ee
   Note that for the construction of the quantized action, the two complex Fields $H_L, H_R$ are replaced by one complex Field $H$, and the complex Field $S$ in the adjoint representation of the SU(5) gauge Theory, is  replaced by a real Field  $K$.
The Old Zinn Sources are obtained from:
\[
\G_{Ri}= \fr{\d \cG}{\d H_L^i}=
\fr{1}{\sqrt{2}}
 \lt ( 
\G_i + i \h_i 
\rt)
;\; 
\G_{L}^{i}= \fr{\d \cG}{\d H_{Ri}}=
\fr{1}{\sqrt{2}}
 \lt ( 
\ov \G^i + i \ov \h^i 
\rt)
;
\]
\[
{\ov \G}_{R}^{i}
= \fr{\d \cG}{\d \oH_{Li}}=
\fr{1}{\sqrt{2}}
 \lt ( 
\ov \G^i - i \ov \h^i 
\rt)
;
\ov \G_{L i}= \fr{\d \cG}{\d \oH_{R}^{i}}
=
\fr{1}{\sqrt{2}}
 \lt ( 
 \G_i - i  \h_i 
\rt)
;\]\be
\; \G^a_{S } 
= \fr{\d \cG}{\d   S^a}=
 \fr{1}{\sqrt{2}}\lt(
 \G_K^a 
 +  i \h_K^a 
\rt )
;
\; \ov\G^a_{S } 
= \fr{\d \cG}{\d   \ov S^a}=
 \fr{1}{\sqrt{2}}\lt(
  \G_K^a 
 -  i \h_K^a 
\rt )
\ee

\refstepcounter{orange}
{\bf \theorange}.\;
{\bf Replacement of Old Variables with New Variables:}  
To get the New Action, we take the Old Action in Equation (\ref{oldaction}),
and replace the Old variables with the New ones.  
For example:
\[
\cA_{\rm Old;\; Zinn\; L} = \int d^4 x \lt \{
i \G_{Ri}
 T^{ai}_{j } H_L^j
\w^a 
+ i \ov \G_{Li}
 T^{ai}_{j} \oH_{R}^{j}
\w^a \rt \}
\]\[
\ra
 \int d^4 x \lt \{
i 
\fr{1}{\sqrt{2}}
 \lt ( 
\G_i + i \h_i 
\rt)
 T^{ai}_{\;\;\;j }
 \fr{1}{\sqrt{2}}
\lt (H^j -i J^j\rt )
\w^a 
\rt \}
\]
\be
+i 
\fr{1}{\sqrt{2}}
 \lt ( 
 \G_i - i  \h_i 
\rt)
T^{ai}_{\;\;\;j } 
\fr{1}{\sqrt{2}}
\lt (H^j +i  J^j\rt )
\w^a 
= \cA_{\rm New;\; Zinn\; L} = 
i  \int d^4 x \lt \{
\G_i   
 T^{ai}_{\;\;\;j }
H^j  
\w^a 
 + 
    \h_i 
T^{ai}_{\;\;\;j } 
  J^j
\w^a 
\rt \}
\ee 
Similarly, consider the following part of the Old Action from  (\ref{cLLold}): 
\be
{\cL}_L +{\cL}_R =
 -
 D_{ \m j}^{i} 
H_{L}^{j} 
 \ovD_i^{\m k} 
\oH_{Lk} 
  -
\ovD_{k}^{\m j} 
\oH_{R}^{k}   
 D_{ \m j}^{i} 
H_{Ri} 
\Ra
\eb
 -
 D_{ \m j}^{i} 
\fr{1}{\sqrt{2}}
\lt (H^j -i J^j\rt )
 \ovD_{i}^{\m k} 
 \fr{1}{\sqrt{2}}
\lt (\oH_k +i \oJ_k\rt )
\eb  -
D_{k}^{\m j} 
\fr{1}{\sqrt{2}}
\lt (H^k +i  J^k\rt )
 \ovD_{ \m j}^{i} 
\fr{1}{\sqrt{2}}
\lt (\oH_i -i \oJ_i\rt )
\ee
and this is
\be
{\cL}_L +{\cL}_R \Ra
 -
 D_{ \m j}^{i} H^j 
 \ovD_{i}^{\m k} 
 \oH_k 
 -
 D_{ \m j}^{i} J^j 
 \ovD_{i}^{\m k} 
 \oJ_k 
\ee
Similarly
\be {\cL}_S 
=-
D^{ab}_{\m} S^b
D^{\m ad} \oS^d 
\Ra
-
D_{\m}^{ab}
 K^b
D_{\m}^{ad}
 K^d
-
D_{\m}^{ab}
 J_K^b
D_{\m}^{ad}
 J_K^d
\ee

\refstepcounter{orange}
{\bf \theorange}.\;
{\bf The New Action has only 34 Real Scalars:}  
\la{newFieldaction}
Now let us collect the New Action together. 
The following two pieces are unchanged:
\be
\cA_{\rm New; \; Gauge} = \int d^4 x \lt \{
\fr{-1}{4} G^a_{\m\n} G^{a \m\n}
\rt \}
\la{newgauge} 
\ee
\be
\cA_{\rm New; \; Ghost,\; Gauge\; Fixing,\; and\;Antighost\; Zinn}
= \int d^4 x  \lt \{
-
\fr{1}{2 \a} 
\lt ( \pa_{\m}V^{\m a} - U^a 
\rt)^2
- \x^a 
 \lt ( 
 \pa_{\m}
D^{\m ab}\w^b
\rt )
\rt \}
\la{newggf}
\ee
Then we get kinetic actions for only 10+ 24 =34 real scalars, whereas there were 68 real scalars in (\ref{oldkinetic}):
\be
\cA_{\rm New; \; Higgs\;Kinetic\; Fields} = \int d^4 x \lt \{
 -
 D_{ \m j}^{i} H^j 
 \ovD_{i}^{\m k} 
 \oH_k 
 -
D_{\m}^{ab}
 K^b
D_{\m}^{ad}
 K^d
\rt \}
\la{newkinetic} 
\ee
The Zinn part that arises from the old kinetic scalar actions contains the other 34 scalars, but now they are Zinn sources rather than physical scalar fields:
\be
\cA_{\rm New; \; Higgs\;Kinetic\; Source \;Terms\; Zinn} = \int d^4 x \lt \{
 -
 D_{ \m j}^{i} J^j 
 \ovD_{i}^{\m k} 
 \oJ_k 
-
D_{\m}^{ab}
 J_K^b
D^{\m ad}
 J_K^d
\rt \}
\la{newzinnkin}
\ee
The Zinn `rotation type' action now reduces to:
\[
\cA_{\rm New\; Zinn} = \int d^4 x \lt \{
i   \G_i 
T^{ai}_{j } 
 H^j
\w^a 
+
i   \h_i 
T^{ai}_{j } 
 J^j
\w^a 
-
i \ov \G^j 
T^{ai}_{j } 
\oH_i  
\w^a 
-
i \ov \h^j 
T^{ai}_{j } 
\oJ_i  
\w^a 
\rt.
\]\be
\lt.
+
 \G_K^a 
f^{abc} 
K^b 
\w^c 
+
 \h_K^a 
f^{abc} 
J_K^b 
\w^c 
+
\S^{a\m}
D_{ \m}^{ab}
 \w^b
- \fr{1}{2} M^{a}
f^{abc} \w^b
 \w^c
\rt\}
\la{newzinn}
\ee
We will not try to follow what happens with the Potential terms in great detail here. We assume for now that the VEVs are zero. The pure Field part is
\[
\cA_{\rm New; \;  Higgs\; Potential} = \]\be
\int d^4 x \lt \{
\cP\lt [H_L \ra \fr{1}{\sqrt{2}}H,H_R\ra \fr{1}{\sqrt{2}}\oH, S\ra \fr{1}{\sqrt{2}}K, \oH_L\ra \fr{1}{\sqrt{2}}\oH,\oH_R\ra \fr{1}{\sqrt{2}}H, \oS \ra \fr{1}{\sqrt{2}}K\rt]
\rt \}
\la{newhigspot}
\ee
and the part that has at least one Source is:
\be
\cA_{\rm New; \;  Potential\;with \;Sources} =(\ref{oldhigspot})\;  ( {\rm after \;the\;substitutions 
 \; (\ref{fsubs2})}) \;\;{\rm minus} \;\; (\ref{newhigspot})
\la{newhigszinn}
\ee
So the New Action $\cA_{\rm New}= \int d^4 x {\cal L}_{\rm New}$ consists of the sum:
\be
\cA_{\rm New}= \int d^4 x {\cal L}_{\rm New}
= (\ref{newgauge})+
(\ref{newggf})+
(\ref{newkinetic})
+
(\ref{newzinnkin})+ 
(\ref{newzinn})+
(\ref{newhigspot})
+(\ref{newhigszinn})
\la{newaction}
\ee

\refstepcounter{orange}
{\bf \theorange}.\;
{\bf The Old and  New Actions both satisfy their Master Equations:}\;
\la{mastermap}
Recall that
 $\cA_{\rm Old}$, which is defined by (\ref{oldaction}), satisfies the Old Master Equation 
\be
\cM_{\rm Old}[\cA_{\rm Old} ] 
=0
\ee
where $\cM_{\rm Old}$ is defined by (\ref{meold}).
Because we obtain $\cA_{\rm New}$, in  (\ref{newaction}),  from $\cA_{\rm Old}$, 
using the Exchange Transformation generated by 
$\cG$ in (\ref{calG}),  it follows that $\cA_{\rm New}$  must satisfy the transformed Master Equation, which is:
\be
\cM_{\rm New}[\cA_{\rm New} ] =0
\la{newmaseqforac}
\ee
where\footnote{The * indicates the complex conjugates of the two previous terms.  The last five terms are real.}
\[
\cM_{\rm New}[{\cal X}] 
= \int d^4 x \lt \{ 
\fr{\d {\cal X}}{\d H^i}
\fr{\d {\cal X}}{\d \G_{i}}
+
\fr{\d {\cal X}}{\d \h^i}
\fr{\d {\cal X}}{\d J_{i}}
+ *
+\fr{\d {\cal X}}{\d K^a}
\fr{\d {\cal X}}{\d \G^a}
+\fr{\d {\cal X}}{\d \h^a}
\fr{\d {\cal X}}{\d J^a}
\rt.\]
\be
\lt.+
\fr{\d {\cal X}}{\d V^a_{\m}}
\fr{\d {\cal X}}{\d \S^{a \m}}
+
\fr{\d {\cal X}}{\d M^a}
\fr{\d {\cal X}}{\d \w^a}
+
\fr{\d {\cal X}}{\d U^a}
\fr{\d {\cal X}}{\d \x^a}
\rt \}
\la{menew}
\ee

\refstepcounter{orange}
{\bf \theorange}.\;
{\bf The New Theory:}\;
\la{newpathintegral}
The New Theory is defined using the New Action, which is just the Old Action expressed in the New Variables.  But there is an interesting twist here.  We must be careful to integrate over the New Fields as integration variables.  So here is the definition of $Z_{\rm New}$:
\be
Z_{\rm New}\lt [ j^{ k},  j_{K}^{a} , j^a_{V, \m} ,
\z^a_{\w } , \z^a_{\x }; 
 \G_{i}
, \G_{K}^a
, \S^{a\m}
, M^{a},J^i,J_K^a
,
U^a
\rt ]
=
\int d ({\rm New \;Fields})\;
 e^{i \int d^4 y \lt \{
{\cL}_{\rm New} +{\cL}_{\rm\;New\; Field\; Sources}\rt \}}
\ee
where we define

\be
d ({\rm New \;Fields})\;
=   \P_{x,i,\m,a}    d H^i   d \oH_{i}   d K^a d \h_{i}   d \ov \h^i  d \h^a  
d V_{\m}^a d \w^a d \x^a 
\la{dnewfields}
\ee
Note that in (\ref{dnewfields}), 
 the New Field variables 
 have replaced the Old field variables    that were in 
(\ref{doldfields}):
\be
     d H_L^i   d \oH_{Li}  d H_{Ri}   d \oH_{R}^i  d S^a d \oS^a 
 \Ra d H^i   d \oH_{i}   d K^a d \h_{i}   d \ov \h^i  d \h^a 
 \ee
These New Field variables are in
\be
{\cL}_{\rm New\;Field\; Sources} = 
\lt \{
j_{k} H^k + {\ov j}_H^{ k} \ov H_{k} + j_{K}^{a} K^a   -\z_{H }^i \h_i  -{\ov \z}_{Hi} {\ov \h}^i 
  - \z_{K}^a \h_K^a
+ j^a_{V,\m} V^{a \m}  -
\z^a_{\w } \w^{a}  -
\z^a_{\x } \x^{a} 
\rt \}
\la{newfieldsources}
\ee

\refstepcounter{orange}
{\bf \theorange}.\;
{\bf The New Invariance:}\;
\la{newinv}
Now, by analogy with Paragraph \ref{oldjacobian} above,
 we take the following change of integration variables ($\varepsilon$ was used above in (\ref{hprime}))
\be
H^i \ra H^i +\D H^i \equiv H^i +\varepsilon \d H^i = H^i + \varepsilon \fr{\d \cA_{\rm New}}{\d \G_{i}}
= H^i  + i  \varepsilon
T^{ai}_{j } 
 H^j
\w^a 
\ee
with the following variations for the other New Fields:
\be
\D  K^a =\varepsilon  \fr{\d \cA_{\rm New}}{\d \G^a}= \varepsilon f^{abc} 
K^b 
\w^c 
\ee

\be
\D  \h_i = \varepsilon \fr{\d \cA_{\rm New}}{\d J^{i}}
=
i  \varepsilon \h_k 
T^{ak}_{i } 
\w^a 
+ 
\varepsilon D_{ \m j}^{i} 
 \ovD_{k}^{\m j} 
\oJ^{k} +  \varepsilon  \fr{\d \cA_{\rm New; \;  Potential\;with \;Sources} 
}{\d J^i} \ee
\be
\D  \h^a =  \varepsilon \fr{\d \cA_{\rm New}}{\d J_K^{a}}=
\varepsilon\h^b
f^{abc} 
\w^c + \varepsilon   D_{\m}^{ab}
D^{\m bd}
 J_K^d
+  \varepsilon  \fr{\d \cA_{\rm New; \;  Potential\;with \;Sources} 
}{\d J_K^a}
\ee

\be
\D  V_{\m}^a = \varepsilon  \fr{\d \cA_{\rm New}}{\d \S^{a\m}}=\varepsilon  D^{ab}_{\m} \w^b
\ee
\be
\D  \w^a = \varepsilon  \fr{\d \cA_{\rm New}}{\d M^a}=- \varepsilon  \fr{1}{2}
f^{abc} \w^b
 \w^c
\ee
\be
\D  \x^a = \varepsilon  \fr{\d \cA_{\rm New}}{\d U^a}= \varepsilon    \fr{1}{ \a} 
\lt ( \pa_{\m}V^{\m a} + U^a 
\rt)
\ee

The New Jacobian Determinant here is again 1.  Consider for example:
\be
{\h'}^a(x)=\h^a(x) +
\varepsilon\h^b(x) 
f^{abc} 
\w^c (x)
+ \varepsilon   D_{\m}^{ab}
D^{\m bd}
 J_K^d(x)
+  \varepsilon  \fr{\d \cA_{\rm New; \;  Potential\;with \;Sources} 
}{\d J_K^a}(x)
\ee
Again, as in Paragraph \ref{oldjacobian}, all the derivatives are non-diagonal except for the one with zero trace.  This suffices to yield a unit Jacobian.

\refstepcounter{orange}
{\bf \theorange}.\;
{\bf Derivation of the New Master Equation for the 1PI Generating Functional:}  
 \la{newmasterfor1PI}
Now we go through the same steps as in Paragraph \ref{oldmasterfor1PI}. 
 Using the above transformation of variables, we see that
\[
\int d^4 x \lt <
j_{k} \fr{\d Z}{\d \G_{k}}  + {\ov j}_H^{ k}  \fr{\d Z}{\d \oG^{k}}  + j_{K}^{a}  \fr{\d Z}{\d \G_K^{a}}  
  +\z_{H }^i \fr{\d Z}{\d J^{i}}  +{\ov \z}_{Hi}  \fr{\d Z}{\d \oJ_{i}} 
  + \z_{K}^a  \fr{\d Z}{\d J_K^{a}}  
\rt.\]
\be
\lt.
+ j^a_{V,\m} \fr{\d Z}{\d \S_{\m}^{a}}   +
\z^a_{\w }  \fr{\d Z}{\d M^{a}} +
\z^a_{\x }  \fr{\d Z}{\d U^a}
\rt >_{\rm New}=0
\la{newwardfromshift}
\ee
We define the Generating Functional $ Z_{\rm New\;C}$ of Connected Green's Functions:
\be
Z_{\rm New}= e^{i Z_{\rm New\;C}}
\ee
and also define the Legendre Transform of the above to be  the Generating Functional $\cG_{\rm New\; 1PI} $ of One Particle Irreducible Green's Functions:
\be
Z_{\rm New\;C}
= \cG_{\rm New\; 1PI} + {\cL}_{\rm New\;Field\; Sources}
\ee
where
\be
\fr{\d Z_{\rm New\;C}}{\d H^i}= \fr{\d \cG_{\rm New\; 1PI}}{\d H^i} +j_{i}=0
\ee
\be
\fr{\d Z_{\rm New\;C}}{\d V_{\m}^a}= \fr{\d \cG_{\rm New\; 1PI}}{\d V_{\m}^a} + j^a_{V,\m} =0
\ee
with similar definitions for the other Sources in expression (\ref{newfieldsources}).  
This implies that
the  New Master Equation has the form
\be
\cM_{\rm New} \lt [\cG_{\rm New\; 1PI}\rt ] =0
\ee
This is also true for the zeroth approximation $\cG_{\rm New\; 1PI}\ra \cA_{\rm New}$, and so we generate Equation (\ref{newmaseqforac}) again.

\refstepcounter{orange}
{\bf \theorange}.\;
{\bf Summary of the Field parts of the Old and New Actions and the suppression from $68 \ra 34$ real scalar bosons:}\;
\la{summary}
The Field parts of the Old and New Actions are different.  The following are in both of them:
\be
\cA_{\rm   Gauge} = 
\int d^4 x \lt \{
\fr{-1}{4} G^a_{\m\n} G^{a \m\n}
\rt \}
\ee
\be
\cA_{\rm  Gauge\; Fixing}
=
\int d^4 x \lt \{
-
\fr{1}{2 \a} 
\lt ( \pa_{\m}V^{\m a} 
\rt)^2
- \x^a 
 \lt ( 
 \pa_{\m}
D^{\m ab}\w^b
\rt )
\rt \}
\ee
However the Old Action and the New Action differ in the kinetic terms for Scalars and in the Scalar Potential terms.  The Field part of the  Old  Action has 10 + 10 + 48 = 68 real scalar bosons:
\be
{\cL}_{\rm Old;\;Scalar\;Kinetic}
 =
 \lt \{
 -
 D_{ \m j}^{i} 
H_{L}^{j} 
 \ovD_i^{\m k} 
\oH_{Lk} 
  -
 \ovD_{ \m j}^{i} 
H_{Ri} 
D_{k}^{\m j} 
\oH_{R}^{k}   
-
D^{ab}_{\m} S^b
D^{\m ad} \oS^d
\rt\} 
\ee
\be
\cA_{\rm Old; \; Higgs\; Potential} = \int d^4 x \lt \{
\cP\lt [H_L,H_R, S, \oH_L,\oH_R, \oS \rt]
\rt \}
\ee
and the Field part of the  New  Action has only $\fr{68}{2}=34=10 + 24$ real scalar bosons:
\be
{\cL}_{\rm New;\;Scalar\;Kinetic}
=
\lt \{
-
 D_{ \m j}^{i} H^j 
 \ovD_{i}^{\m k} 
 \oH_k 
 -
D_{\m}^{ab}
 K^b
D_{\m}^{ad}
 K^d
\rt \}
\ee
\[
\cA_{\rm New; \;  Higgs\; Potential} = \]\be
\int d^4 x \lt \{
\cP\lt [H_L \ra \fr{1}{\sqrt{2}}H,H_R\ra \fr{1}{\sqrt{2}}\oH, S\ra \fr{1}{\sqrt{2}}K, \oH_L\ra \fr{1}{\sqrt{2}}\oH,\oH_R\ra \fr{1}{\sqrt{2}}H, \oS \ra \fr{1}{\sqrt{2}}K\rt]
\rt \}
\ee

Note that the above Field terms for the New Action are just what one would expect if one simply wrote that theory down from the start.  The New Antighost Fields $\h^i, \ov \h_i,\h^a$ and the New J-type Zinn Sources play no role in the physics of the New Action--their role is to keep the symmetry alive through the Zinn Source part.

\refstepcounter{orange}
{\bf \theorange}.\;
{\bf Conclusion and the  \SG\ case:}\;
\la{conclusion}
We have seen that the \ET\  permits one to reduce the Scalar Representations while keeping essentially the same Action, by transforming some of the Old Scalars into New Zinn Scalar Sources, and the conjugate Old Zinn Scalar Sources into New Antighosts.
Clearly we could get the New Theory much more simply by just choosing the $H^i,K^a$ representations to start with, rather than by suppressing the original Old Action with the \ET.  
 However this is not feasible for the \SG\ theory because   SUSY requires Chiral Scalar representations, and these need Scalars $H_L^i,H_{Ri},S^a$ to give masses to the Quarks and Leptons and to break gauge symmetry.
 It is clear that the Exchange Transformation in \ci{su5gust}, which is identical to the one in this paper,  eliminates 34 of the 68 real Scalars, so that SUSY gets  shifted into the $J^i, J_K^a$ sources in \ci{su5gust}, and the SUSY representations are spoiled by the \ET\ in that sense.
Consequently this appears to be a new way to `break SUSY', although really SUSY is still there, but realized through the New Antighost  and New $J$ Zinn Source terms, as above.

In this paper we have not tried to analyze issues that arise from VEVs for the Scalars in the context of Suppressed SUSY.  That important subject is discussed in \ci{su5gust,four,five} however.
When the VEV appears, as it does in \ci{su5gust}, it gives a mass to both the vector bosons and to the gravitino, and clearly this cannot be a theory with an unbroken SUSY spectrum. The theory has the wrong Scalar Field content for that.
Moreover, as was seen in \ci{su5gust}, the VEV of the vacuum remains zero, at least at tree level.
This amounts to spontaneous breaking of gauge symmetry.
But it cannot be spontaneous breaking of supersymmetry, because the Field content is not that of Supersymmetry. 

In order to treat the fermion spectrum in \ci{su5gust} correctly, at tree level, more work is required for the fermions.  There are similar \ET s to be done for the Higgsinos and Gauginos.
 There is probably something like a Goldstino involved, and some careful analysis of the gauge fixing and ghost action for the gravitino is needed.  As far as the author knows this has never been done in a convincing way even for the spontaneously broken theory.  For example, where is the gauge parameter (like $\a$ above in Equation (\ref{laterggf})) for the gravitino gauge fixing? 
We have not yet attempted to write out the \ME\ for the Supergravity Theory in detail. This is a daunting task.     The author's guess, at this time, is that no unsolvable new issue arises, but that needs a demonstration, and completing this task is a major piece of work.  
The \SG\ action also contains a Kahler invariance.  As with SUSY itself, we expect this to remain present, but to be spread into the Zinn sources.
Of course our problem is simpler--we do not require full generality for the Supergravity Master Equation, since we are in a flat known Kahler metric and the kinetic terms of the gauge theory are also simple, and the superpotential is known.

\begin{center}
 {\bf Acknowledgments}
\end{center}
\vspace{.1cm}

  I thank Doug Baxter, Carlo Becchi, Philip Candelas,  James Dodd, Mike Duff, Richard Golding,  Dylan Harries, Chris Hull, Sergei Kuzenko,  Pierre Ramond, Graham Ross, Peter Scharbach,    Kelly Stelle,  Xerxes Tata,  J.C. Taylor, Steven Weinberg,  and Peter West for stimulating correspondence and conversations.

\tiny
\articlenumber\\
\today
\end{document}